\documentclass[12pt]{iopart}
\begin{document}

\title[Author guidelines for IOP journals in  \LaTeXe]{ALICE potential for Heavy-Flavour Physics}

\author{Gin\'es Mart\'{\i}nez Garc\'{\i}a for the ALICE Collaboration}

\address{Subatech (Universit\'e de Nantes, Ecole des Mines and CNRS/IN2P3), Nantes, France}
\ead{Gines.Martinez@subatech.in2p3.fr}
\begin{abstract}
Heavy Quarks  will be abundantly produced in Heavy Ion Collisions at LHC energies.
Both, the production of open heavy flavoured mesons and quarkonia will probe the
strongly interacting medium created in these reactions.
In particular, the ALICE detector will be able to measure heavy flavour production down to low transverse momentum,
combining leptonic and hadronic channels, covering a large rapidity range $|\eta|<0.9$ and 
$-4<\eta<-2.5$.
In this talk we will present the main physics motivations for the study of heavy flavour production 
at LHC energies and some examples of physics analyses developed by the heavy flavour 
working group of ALICE.
\end{abstract}

%\maketitle

\section{Introduction}
The Large Hadron Collider (LHC) will provide ion beams at ultra-relativistic energies. 
Luminosity of lead lead collisions is expected to be $<\mathcal{L}>=5 \cdot 10^{26}$cm$^{-2}$s$^{-1}$ at $\sqrt{s}=$5.5A TeV.
In addition, the LHC will deliver proton proton collisions and proton lead collisions,
providing a solid baseline for the study of medium effects in heavy ion collisions (HIC) 
\cite{AlicePPR1}\footnote{The duration of a proton-proton run at 14 TeV is expected to be $10^7$ s. For the other systems,  the duration of the run is expected to be $10^6$ s.}.
At such ultra-relativistic energies, new observables will become accessible to study the strongly interacting medium
that will be produced in HIC.
In particular, the increase of the production cross-section of hard probes
will allow to probe the hadronic matter via high p$_T$ hadrons, jets, high p$_T$ photons and heavy flavours (HF)
\cite{AlicePPR1,AlicePPR2,Anti06}. 
In this talk we will present the main physics motivations for the study of HF production 
at LHC energies and a few examples of physics analysis developed by the ALICE collaboration.

\section{Heavy Flavour production at LHC energies}
In HIC, heavy quarks are expected to be produced in the first stages of the collision  and   
to coexist with the surrounding medium due to their long life-time.
Therefore HF production will probe the strongly interacting medium produced in these reactions.

\begin{itemize}
\item {\bf Initial production.}
Charm quarks will be copiously produced in hadron collisions at LHC energies via predominantly 
gluon-gluon fusion: 
more than 100 $c\bar{c}$ and around 5 $b\bar{b}$ pairs per central Pb+Pb collision 
are expected \cite{AlicePPR2, Yell03}.
In the ALICE acceptance, HF production will probe the gluon small Bjorken-x domain ($x \in [10^{-5},10^{-3}]$).
In the case of HIC, one should expect that gluon shadowing modifies the heavy quark production 
with respect to the binary scaling.
Indeed, a suppression of about 20\% (10\%) for charm (beauty) in p+Pb 
collisions is expected due to gluon shadowing \cite{Rauf04}.
About 1\% of the heavy quark pairs should result in the formation of colourless bound states: the quarkonia. 
The process of formation of quarkonia from the initially produced $Q\bar{Q}$ pair is a non-perturbative process.
In this respect, quarkonia production is challenging for QCD theory since it must be addressed by non-perturbative 
models like non-relativistic QCD (see \cite{Yell03} and reference therein).

\item {\bf Heavy quarks in hot and dense medium.} 
Heavy quarks will propagate in matter mainly formed by gluons and light quarks ($u,d,s$).
It is still an open question how heavy quarks will behave in such a medium,
although PHENIX and STAR experiments are already addressing this question \cite{Cald07}.
Will heavy quarks thermalize? Will heavy quarks develop collective motion? 
Some models assume that heavy quarks behave like \emph{Brownian} particles (m$_Q>$m$_q$) \cite{Goss04} and
their transverse momentum (p$_T$) and rapidity distributions and azimuthal asymmetries 
will probe the properties of the QGP.
Other models do not address the problem of thermalization 
and directly assume statistical coalescence of heavy quarks \cite{Andr03,Thew01,Gran04}. 

The measurement of the $Q\bar{Q}$ bound states will allow to study the properties of the medium
via the Debye screening \cite{Datt04} and/or the gluon dissociation of quarkonia \cite{Gran04}.
In addition, the charmonium states will also allow to study the kinetic recombination of heavy quarks \cite{Thew01} 
and/or their statistical hadronization \cite{Andr03}.

\item {\bf Secondary production.}
$\psi'$ and $\chi_c$ charmonia resonances will decay into $J/\psi$ with a relative contribution 
to the full direct $J/\psi$ production of about 40\%.
$\Upsilon(1S)$ ($\Upsilon(2S)$)  production will be increased by radiative decay of the $\chi_b$ resonance and 
decays of higher $\Upsilon(nS)$ resonances.
It is expected about 45\% (30\%) of feed-down from other bottomonia resonances for the $\Upsilon(1S)$ ($\Upsilon(2S)$).
In addition, there will also be production of $J/\psi$ due to charmonium decay modes of the $B$ mesons.
At LHC energies the relative contribution of $J/\psi$ ($\psi'$) from B-meson decay should be about 22\% (39\%) \cite{Yell03}.
However, the latter will be evaluated  on a statistical basis via the measurement of open beauty production cross-section 
and via the high precision vertexing of quarkonia decay products on an event-by-event basis.
\end{itemize}

\section{Heavy Flavour Physics in the ALICE detector}
A Large Ion Collider Experiment (ALICE)  has been designed for addressing the  
physics of HIC at LHC energies and will study the production of HF in several channels \cite{AlicePPR1}
in  p+p, p+Pb and Pb+Pb collisions.
In the following, we will present a few examples of physics channels addressed by the heavy flavour working group of ALICE:
\begin{itemize}
\item {\bf Heavy flavoured mesons in the hadronic channels.}
Production of open charmed hadrons will be studied in the ALICE central barrel ($|\eta|<0.9$) 
via their decay into 
hadronic channels like $D^0\rightarrow K^-\pi^+$.
Hadrons from secondary vertices will be tracked by the central tracking system of ALICE: 
the six layers of the silicon Internal Tracking System (ITS) and the Time Projection Chamber (TPC).
The Time Of Flight (TOF) detector will allow for pion/kaon discrimination in the low transverse momentum domain (p$_T<~2$ GeV/c).
Cuts on the separation between the primary vertex and secondary tracks, on the secondary vertex quality and on the pointing angle of the reconstructed mother 
particle will allow for improving the signal to background ratio. 
In particular, during one year of data taking at LHC, $D^0$ meson will be studied in the transverse
momentum range from 1 to 20 GeV/c in the $K^-\pi^+$ channel 
in p+p p+Pb and Pb+Pb collisions with a statistical error better than 20\% \cite{AlicePPR2,Dain02, Dain04,Dain06}.
The $D^\pm$ can also be studied by the ALICE detector \cite{Brun06}.
Studies of $D_s$ and $\Lambda_c$ are also in progress.

\item {\bf Beauty detection in the electron channel.}
Open beauty production can be studied via the electron detection in the central barrel.
Electrons will be identified using the information of the Transition Radiation Detector (TRD) 
and their energy loss in the TPC and will be tracked by the ITS, TPC and TRD detectors. 
High precision vertexing will be necessary to select electrons (p$_T>1$GeV/c) created
several hundreds of micrometers away from the primary vertex.
These electrons are mainly produced by semi-electronic decays of charm and beauty hadrons.
The direct measurement of the charmed meson production will allow to evaluate
the contribution of charm to the electron spectrum \cite{AlicePPR2,Anti06b} 
in order to extract the beauty yields.

\item {\bf Beauty and W$^\pm$ detection in the muon and di-muon channels.}
Open beauty production will be studied via the measurement of muon and di-muon production in the muon spectrometer of ALICE 
($-4.0<\eta<-2.5$). 
High p$_T$ muon events will be efficiently triggered by the trigger system of the muon spectrometer \cite{Gueri06}. 
The analysis of the single muons and the correlated continuum of di-muon invariant masses
will allow for studying the beauty production in the p$_T$ range from 1 to 20 GeV/c \cite{AlicePPR2, Guer05, Guer05b, Croc05}.
For p$_T$ higher than 20 GeV/c, the W muon decay will be the dominant contribution to the muon p$_T$ distribution \cite{Cone06, Cone06b}.
W boson production will be studied by measuring the muon p$_T$ distribution in the range from 25 to 50 GeV/c.
W bosons will interact weakly with the QGP and would provide a baseline to study medium induced effects on beauty flavour production.

\item {\bf Quarkonia detection in the di-electron channel.}
Quarkonia production at mid-rapidity ($|y|<0.9$) will be studied in the di-electron channel.
Electrons will be identified by the TRD and TPC, and tracked by the ITS, TPC and TRD.
The invariant mass resolution is expected to be around 30 (90) MeV/c$^2$ for charmonia (bottomonia).
Good statistics of $J/\psi$ events  will be collected for the most central Pb+Pb collisions, more than a hundred thousand $J/\psi$'s 
(and about 900 $\Upsilon(1S)$ events) during one year of data taking \cite{AlicePPR2,Somm07}. 
The high precision vertexing of the ITS system should allow for tagging $J/\psi$ from beauty hadron decays.

\item {\bf Quarkonia detection in the di-muon channel.}
Quarkonia production at rapidities $-4.0<|y|<-2.5$ will also be studied in the di-muon channel.
%Muon events triggered by the trigger system of the muon spectrometer will be tracked along the muon magnet by the tracking system 
%of the muon spectrometer. 
The invariant mass resolution will be around 70 (100) MeV/c$^2$ for charmonia (bottomonia) in central Pb+Pb collisions.
Novel observables like the relative yield of $\Upsilon$'s $N_{\Upsilon(2S)}/N_{\Upsilon(1S)}$ will be studied \cite{AlicePPR2,Dumo05}.
More than half a million $J/\psi$'s are expected to be detected during one Pb+Pb run at LHC.
This will allow to study the charmonia production in the p$_T$ range from 0 to 20 GeV/c.
Quarkonia will also be studied in proton proton collisions, where about 3 millions $J/\psi$ events per run 
will be collected \cite{Stoc07}.
Such a high statistics will open the possibility to study the $J/\psi$ polarization at high transverse momentum \cite{Arna06}.
\end{itemize}

\section{Conclusions}
HF production at LHC energies is a very promising probe to study the properties of the high energy density medium
formed in HIC. The ALICE experiment will address the richness of HF physics in a large rapidity window,
down to very low p$_T$'s  by combining the analysis of hadronic and leptonic channels.
~\\
%\section{References}

\end{document}